\documentclass[twocolumn]{aastex631}

\usepackage{xcolor}

\shorttitle{SED Fitting Systematics}
\shortauthors{Zine \& Salim}

\graphicspath{{./}{figures/}}

\begin{document}

\title{Systematics in the SED Fitting Parameter Estimation of Composite Galaxies}

\author{Katherine Zine}
\affiliation{Department of Astronomy, Indiana University, Bloomington, IN 47405, USA}

\author[0000-0003-2342-7501]{Samir Salim}
\affiliation{Department of Astronomy, Indiana University, Bloomington, IN 47405, USA}

\begin{abstract}

Derivation of physical properties of galaxies using spectral energy distribution (SED) fitting is a powerful method, but can suffer from various systematics arising from model assumptions. Previously, such biases were mostly studied in the context of individual galaxies. In this study, we investigate potential biases arising from performing the SED fitting on the combined light of two galaxies, as would be the case in post-merger systems. We use GALEX-SDSS-WISE Legacy Catalog (GSWLC) of $z<0.3$ galaxies to identify 9,000 galaxy pairs that could eventually merge. For these we investigate if the UV/optical SED fitting accurately determines the stellar mass and (specific) star formation rate if the pair was unresolved (merged). The sum of the stellar masses (and SFRs) of individual galaxies in the pair establishes the ground truth for these quantities. For star forming galaxies no biases ($<$0.1 dex) are found in the stellar mass, SFR, or sSFRs. Moderate systematics in SFR ($\sim$0.1 dex) are found for systems with an extreme contrast in dust content between the two galaxies. We conclude that biases that would arise in the determination of masses and SFRs of post-merger systems on account of the two original galaxies having potentially very different star formation histories and different dust properties are small and that the approach with simple two-component star formation histories is adequate. The approach presented in this study, using flux compositing with empirically determined ground truth, offers new opportunities for testing the results of SED fitting in general.
\end{abstract}



\section{Introduction} \label{sec:intro}
Galaxy mergers play an important role in galaxy evolution. They can lead to dramatic changes in galaxy structure and morphology. For example, when simulating the life of a galaxy undergoing a merger, the galaxies that start off as disk galaxies can become spheroids after a merger \citep[e.g.,][]{{2009ApJ...696..348W}}. Mergers can also augment star formation efficiency and produce starbursts \citep[e.g.,][]{{2011EAS....51..107B}}. The study of galaxy evolution requires unbiased determination of galaxy properties, such as the stellar mass and the star formation rate (SFR). This is potentially more challenging for galaxies that have recently undergone a merger for  two reasons. One is that the merger itself can produce a burst of star formation and therefore the star formation history (SFH) of the merger, especially a recent one, is more complex than those of the pre-merger galaxies. The second arises from the fact that mergers can result from the combination of different types of galaxies, each with its own distinct SFHs and dust contents prior to merging. The goal of this paper is to shed light on the second effect, which we refer to as ``compositing". 

Spectral energy distribution (SED) fitting is an important tool in the study of galaxy evolution due to its ability to simultaneously determine many properties of a galaxy, such as star formation history and stellar mass, based on its light output \citep{{2013ARA&A..51..393C}}. In the context of SED fitting, the SEDs are crude spectra of galaxies consisting of flux measurements at different wavelengths, usually spanning from rest-frame UV to near IR \citep{{2011Ap&SS.331....1W}}. In SED fitting a large number of models are created by specifying parameters relating to SFH, stellar populations, and dust attenuation \citep[hereafter S16]{2016ApJS..227....2S}. The observed SED is compared to model SEDs to find the best fit, or more generally, to build the probability distribution functions for different properties \citep{{2011Ap&SS.331....1W}}. The models used in SED fitting are based on stellar population synthesis, which uses stellar evolution theory and libraries of observed stars of different spectral types or model stars with different physical properties \citep{{2013ARA&A..51..393C}}.

Whereas the observed SEDs are the result of an underlying physics that is being modeled, the models have limitations or simplifications that may affect the accuracy of parameters derived from the SED fitting. Furthermore, the observed SEDs, which have some finite precision, may not be able to accurately distinguish between different models due to the potential degeneracies. Various systematics involved with SED fitting have been studied previously, but usually for the general population of galaxies, which is not likely to have undergone major merging events in the recent past. For example, for the stellar mass of galaxies, \citet{{2015MNRAS.452..235S},{2018MNRAS.476.1532S}}, following the approach of \citet{{2009MNRAS.400.1181Z}}, found that in the case of photometry that was spatially unresolved, the mass found using SED fitting was underestimating the true mass. \citet{{2006ApJ...652...97V}} found that there can be systematics at different redshifts depending on various factors, such as the stellar population synthesis model employed. \citet{{2013MNRAS.435...87M}} found systematics in stellar mass caused by using only one metallicity value and depending on how mass recycling was treated. Systematics were also found by \citet{{2014arXiv1404.0402S}} in colors and mass-to-light ratios when using an exponentially declining model for star formation history. S16 has discussed how two-component star formation histories, where one component is posited to be very old, are able to recover larger stellar masses and remove or reduce the outshining issue \citep{{2014A&A...571A..75M}}. Yet, another set of systematics emerges from the treatment of dust. For example, the use of fixed dust attenuation curves can lead to biased SFRs and stellar masses, especially if the assumed law differs from the average law of observed galaxies \citep{{2020ARA&A..58..529S},{2017MNRAS.472.1372L}}.

Additional sytematics my be present in merging or recently merged systems. \citet{{2009ApJ...696..348W}} studied how systematics may arise during different phases of a merger. They created a simulation that progressed over time and found that during the period of enhanced star formation due to a merger the age and mass were being most underestimated, whereas the underestimation of SFR varied at different points during the merger \citep{{2009ApJ...696..348W}}. \citet{{2015MNRAS.446.1512H}} applied MAGPHYS SED fitting code to hydrodynamical simulations of merging galaxies and found good fits and that most of the parameters were recovered correctly, with the exception of the dust mass, which may have had systematics at various times, when the galaxy was not at near-coalescence. \citet{2014ApJ...785...39L} found that SEDs cannot usually be used to indicate the interactions stage of galaxies, but this does not necessarily mean that the SEDs of merged galaxies are not at least somewhat distinct from the galaxies that have not experienced a merger (at least recently) and that such distinct SEDs would still be well modeled in the course of SED fitting.

Previous studies essentially focused on the systematics arising from the changes produced by the merger itself. However, when two galaxies merge such that we can no longer resolve them individually and therefore cannot separate the SEDs of the original galaxies, an additional set of systematics may arise in the SED fitting due to the fact that we are trying to model the combined light of the galaxies, which is a composite of two galaxies with distinct SFH and different dust properties. The sheer number of possible combinations precludes such composite SFHs or complex dust scenarios to be explicitly included in the library of models, so the question we deal with in the current work is whether the commonly used SFH parameterizations and simple dust models can still provide unbiased results in the case of composite SEDs.

The paper is organized as follows. In Section \ref{sec:data} the data and sample used for this study are described. Section \ref{sec:SED fitting} describes methodology used to assess biases in the SED fitting of composite galaxies. The results are presented in Section \ref{sec:results} and a summary and a discussion of the results is in Section \ref{sec:discussion}.

\section{Data and Sample} \label{sec:data}

The parent sample was taken from the GSWLC-D catalogue (S16). GSWLC is a catalog of galaxy parameters obtained from SED fitting, and consists of galaxies contained in the SDSS spectroscopic sample (redshifts between 0.01 and 0.30) that had ultraviolet (UV) coverage from GALEX (S16). UV observations are necessary to constrain dust-corrected SFRs. The D catalog includes galaxies with the longest exposure times in the UV (longer than 4000 seconds) and therefore has the highest precision of SFR estimates (S16). The catalog contains position in the sky, SDSS identifications, and physical parameters such as the stellar mass, dust attenuation, and star formation rates of 48,401 galaxies (S16). GSWLC is based on the photometry in the FUV, NUV, $u$, $g$, $r$, $i$, and $z$ bands, which will be used in the SED fitting in this work as well.

\begin{figure}[ht!]
\epsscale{1.3}
\plotone{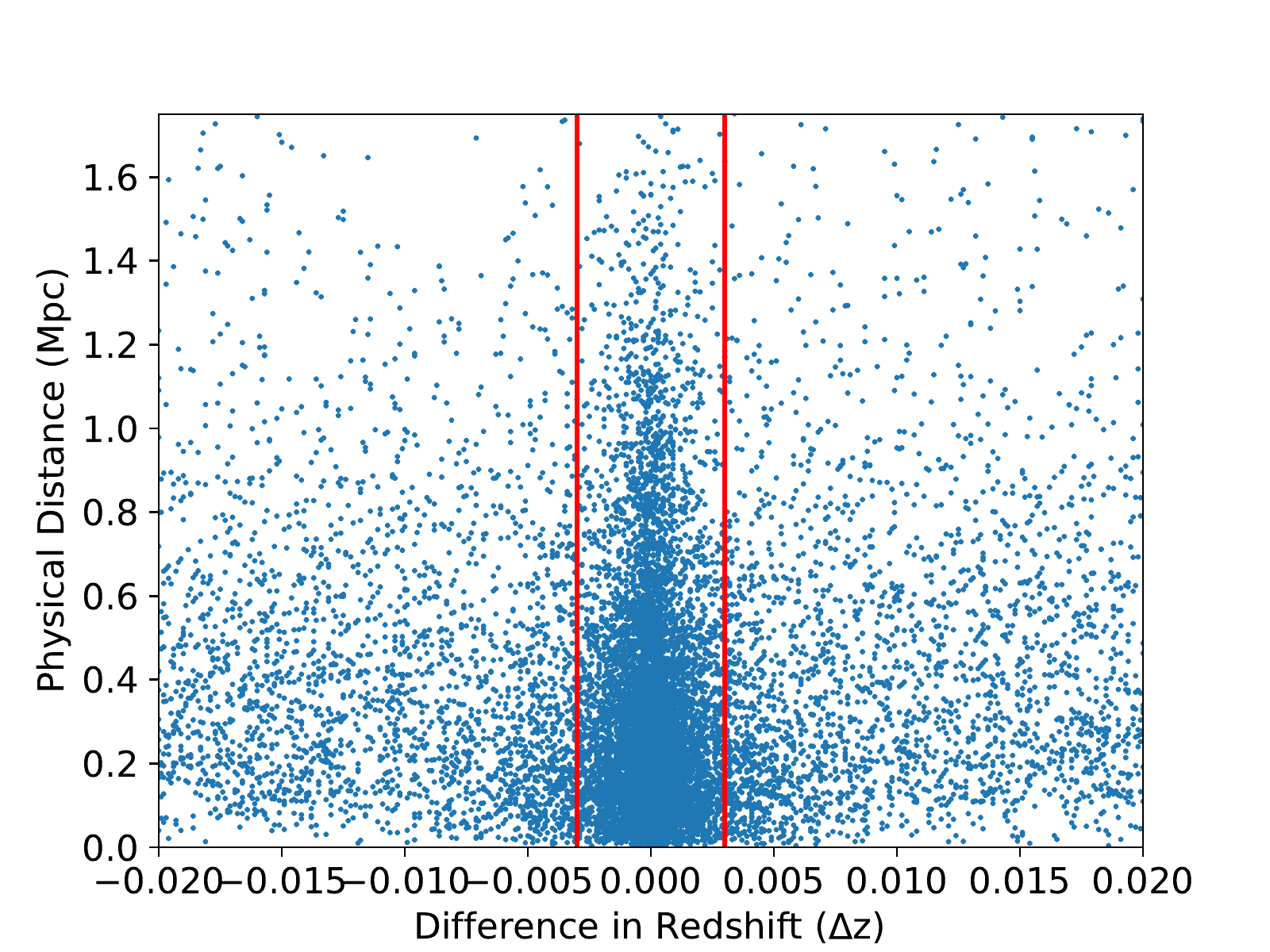}
\caption{Projected physical and redshift separations of galaxies in candidate pairs. The physical separation in Mpc for the two galaxies in the 33,557 pairs plotted against their difference in redshift. Only galaxy pairs between the two red lines were decided to be close enough to be considered potential future mergers and were kept in the analysis.
\label{fig:redshift}}
\end{figure}

One can in principle test biases emerging from the compositing of SEDs by performing the SED fitting on any two galaxies from the parent sample and combining their light. However, in order to make our assessment more pertinent to a physical situation when this compositing may occur, we focus on physical galaxy pairs-galaxies that are close enough to each other that they could realistically merge. The added benefit of such an approach is that each galaxy in the composite will lie at the same distance, so combining their light can be accomplished by simply adding the fluxes.

We perform galaxy pair selection using a two-step process. First, we identify 33,557 candidate pairs, which are defined as a pair with the smallest angular separation from each other not exceeding one degree. In order to establish if the candidate pair is physically associated, we examine their redshifts. In Figure \ref{fig:redshift} we show the physical projected separation of candidate pairs where the galaxies have similar redshifts. We see an overdensity of pairs around $\Delta z=0$. Informed by the appearance of this distribution, for our final sample we require the difference in redshifts to be less than 0.003, corresponding to a maximum radial velocity difference of 900 km $s^{-1}$. Such pairs are all within about 2 Mpc projected distance (similar to the size of galaxy groups), and often much closer, so we do not impose any upper limits on the projected separation. The final sample consists of 9,032 pairs with 15,736 unique galaxies, or 32.5\% of the parent sample. The number of unique galaxies is somewhat smaller than twice the number of pairs because a given galaxy can be a member of more than one pair.

\begin{figure*}[ht!]
    \includegraphics[width=0.5\textwidth]{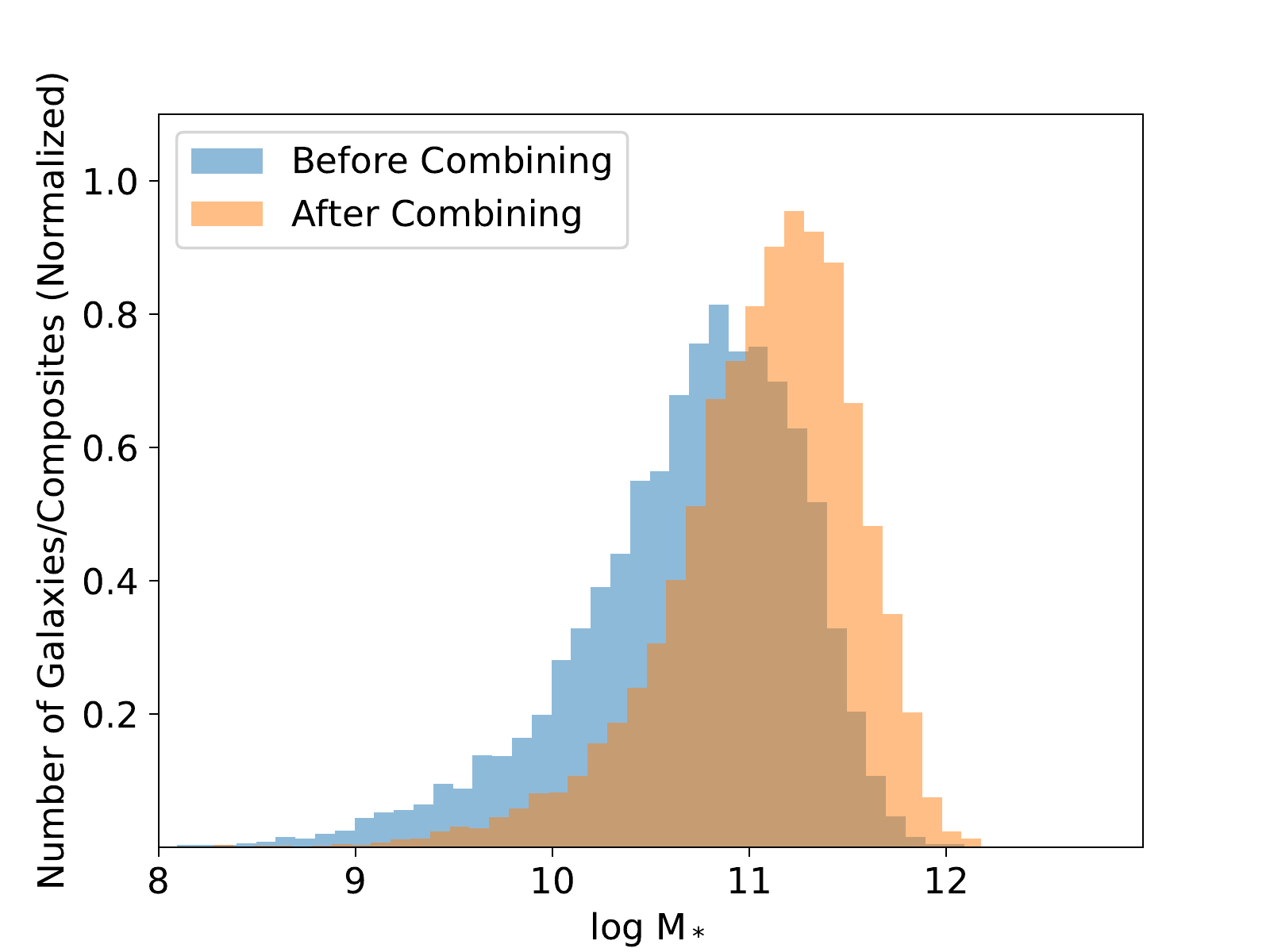}
    \includegraphics[width=0.5\textwidth]{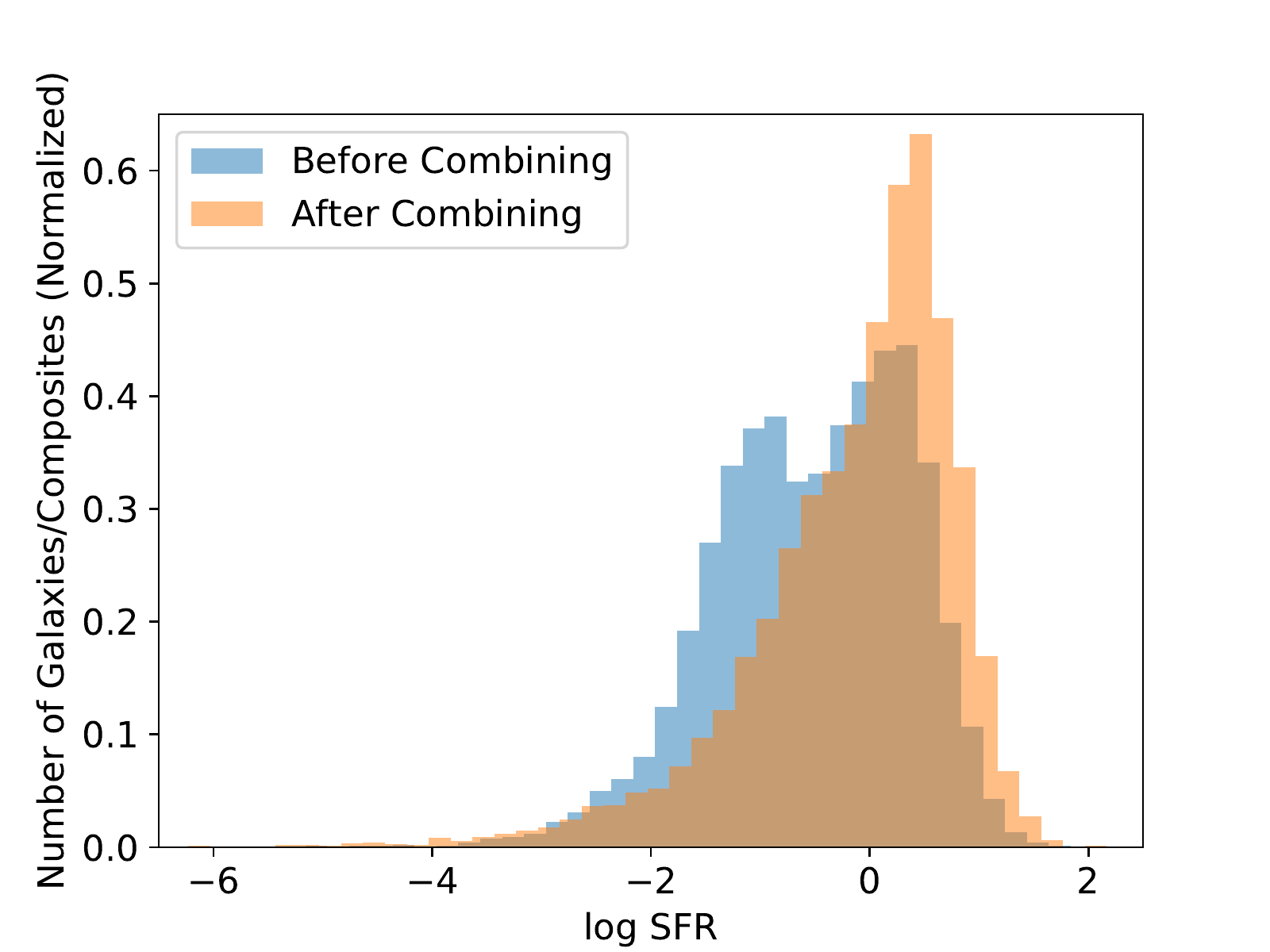}\\
    \includegraphics[width=0.5\textwidth]{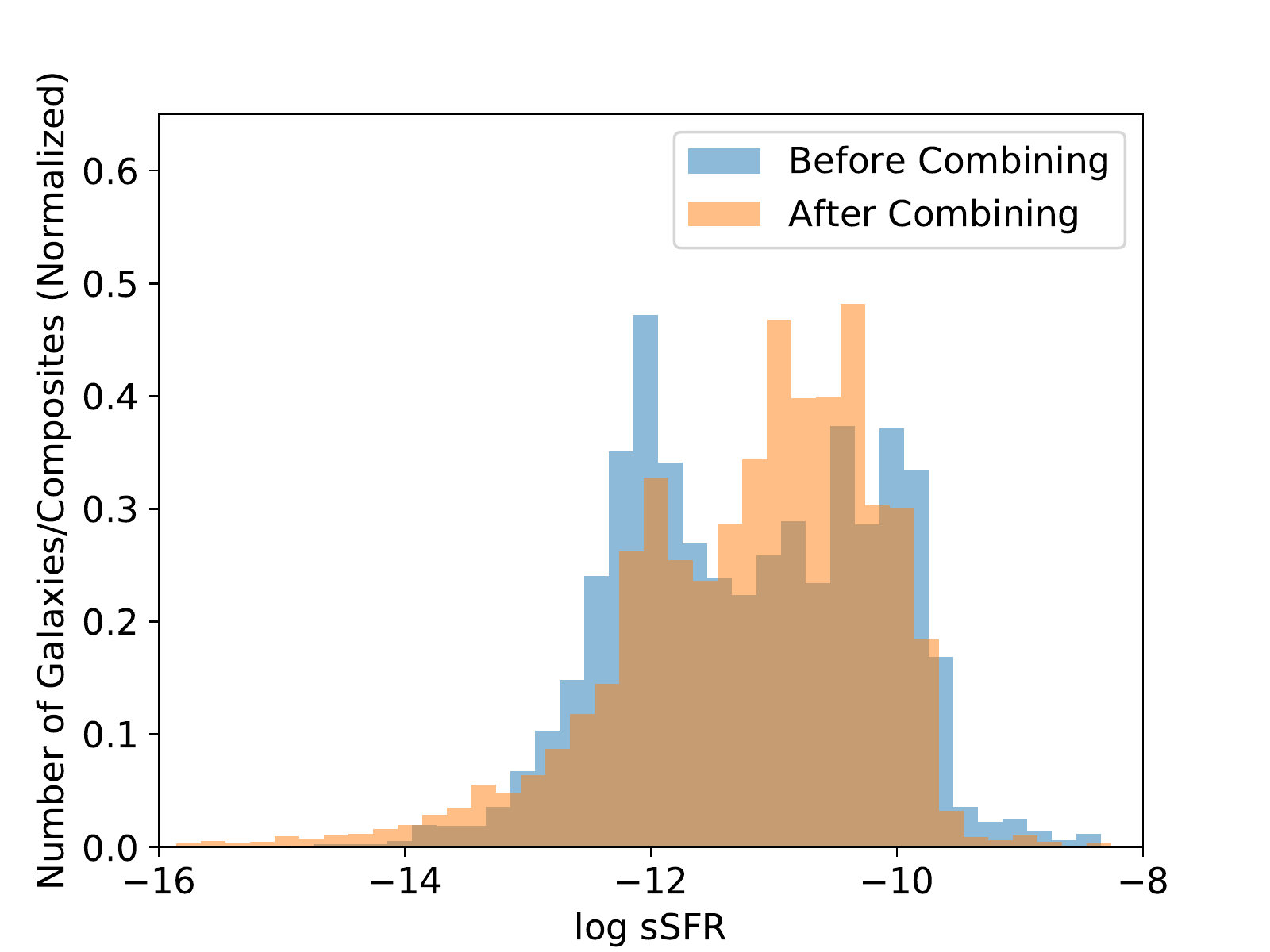}
    \includegraphics[width=0.5\textwidth]{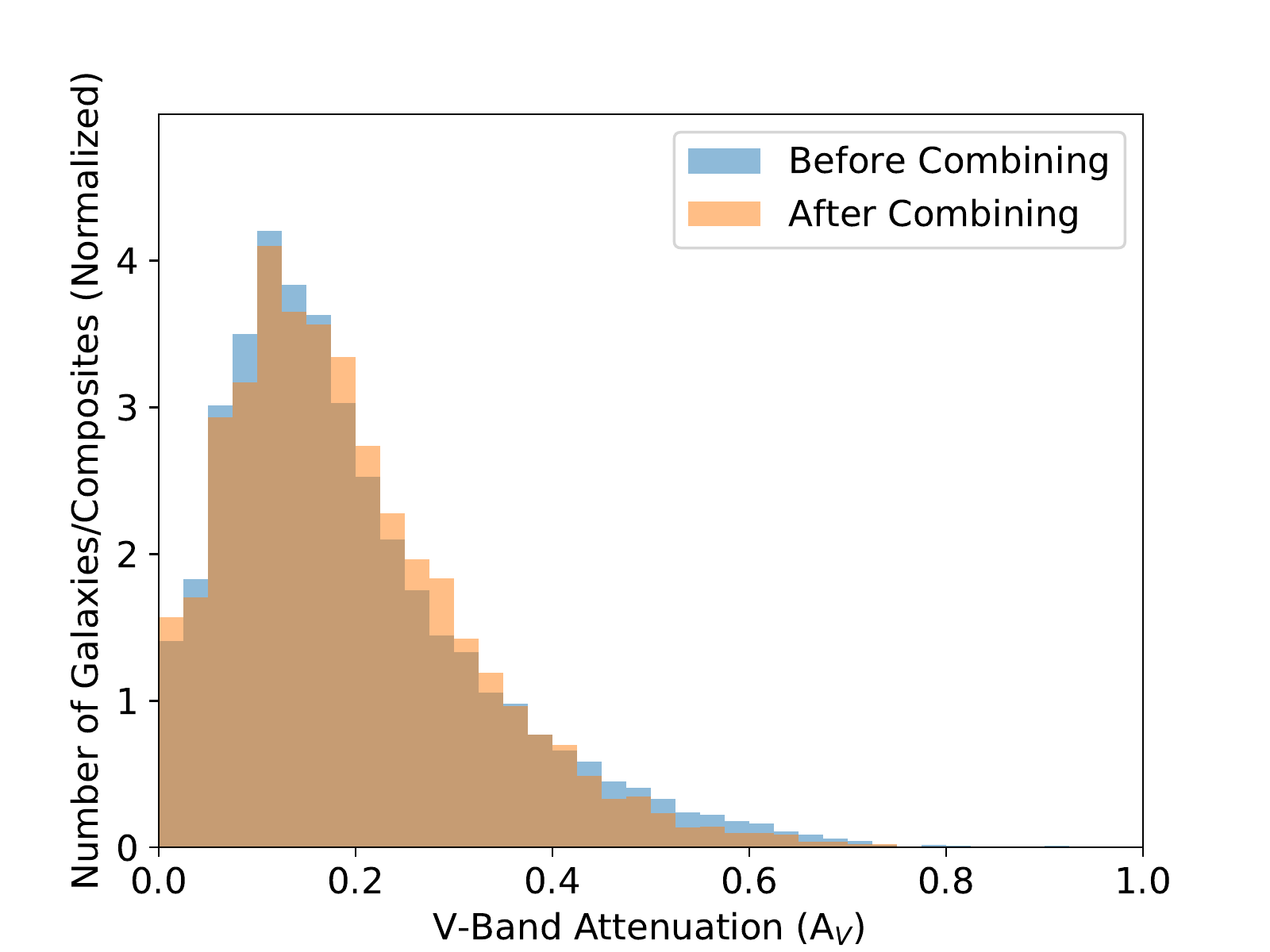}
    \caption{Distributions of galaxy properties in our sample of physical pairs (candidate mergers). Blue histograms show properties of individual galaxies before combining their light, and the peach histograms are the properties of composite galaxies (after their SEDs were combined). Properties include the stellar mass, star formation rate, specific star formation rate, and \emph{V}-band attenuation. \label{fig:1D_hist}}
\end{figure*}

\begin{figure}[ht!]
    \includegraphics[width=\columnwidth]{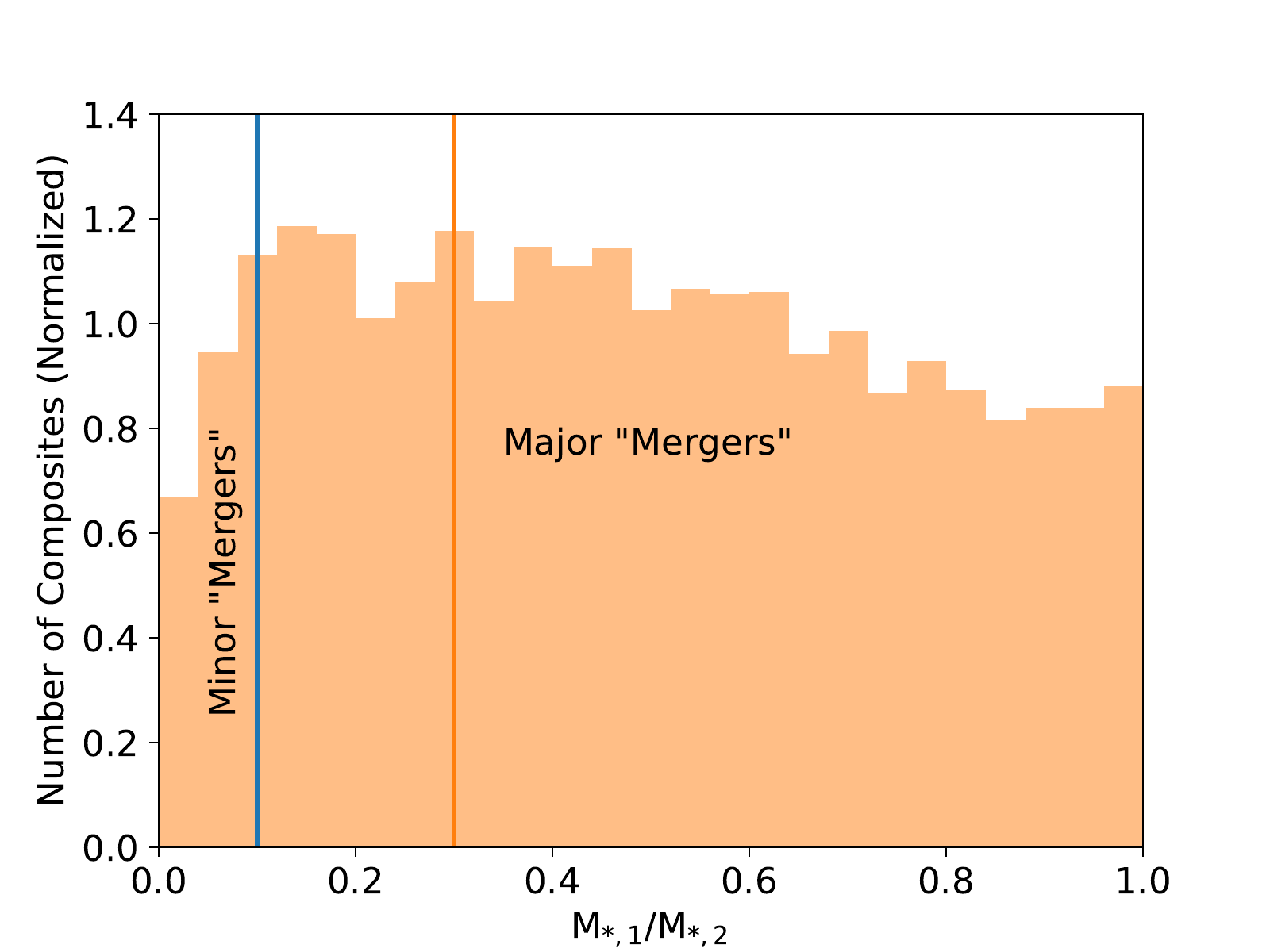}
    \caption{Stellar mass ratio of galaxies making the physical pairs  (candidate mergers). There is a wide range of mass ratios. The majority of pairs have mass ratios that would correspond to a major merger.\label{fig:mass_ratio}}
\end{figure}

\begin{figure}[ht!]
    \includegraphics[width=\columnwidth]{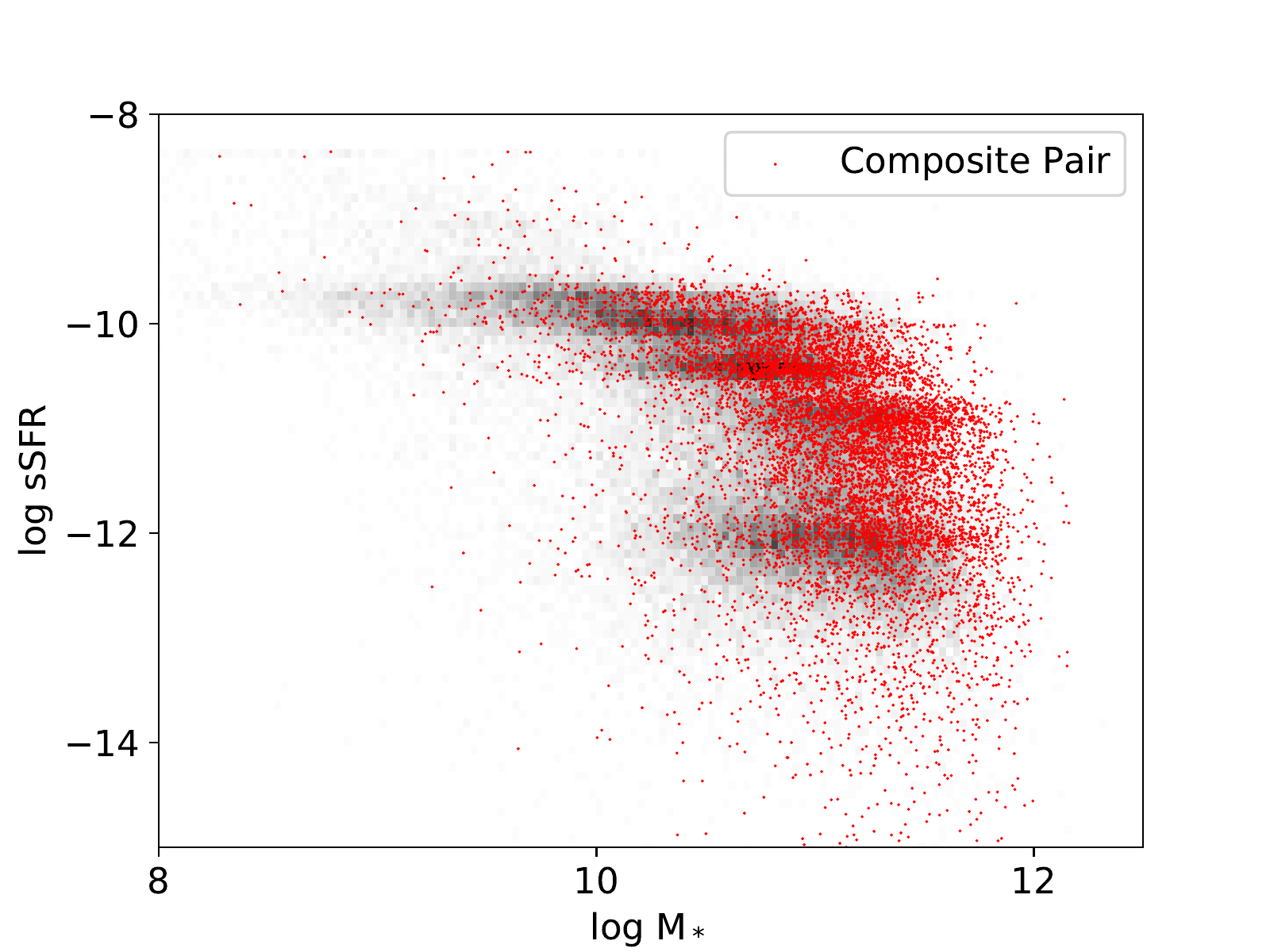}
    \caption{Stellar mass versus specific star formation rate. The grey plot is for all 48,401 galaxies in the parent sample, whereas the composite galaxies are in red. Composites occupy an entire parameter space, but are somewhat less present among the low-mass galaxies.\label{fig:M_SSFR_hist}}
\end{figure}

\section{SED Fitting} \label{sec:SED fitting}
We perform the SED fitting on the galaxies in the parent sample (which includes the individual galaxies from the 9,032 pairs) as well as on the composite galaxies, for which we combine the fluxes of members in each pair prior to SED fitting. 

SED fitting was performed with the 2020 version of the Code Investigating GALaxy Emission, or CIGALE \citep{{2019A&A...622A.103B}}. We used \citet{{2003MNRAS.344.1000B}} stellar population synthesis models calculated for four stellar metallicities and for the Chabrier IMF. Nebular emission lines are included in the models as described in \citet{{2018ApJ...859...11S}}. Importantly, to model the dust, we follow the approach from S16 based on having more than one attenuation curve in the model library. In particular, we use the \citet{{2009A&A...507.1793N}} modification of the \citet{{2000ApJ...533..682C}} attenuation curve, and make models using two slopes each steeper than the original Calzetti curve. To this curve we add the UV bump, in several increments up to twice the strength of the UV bump in the MW. Note that this SED fitting does not include IR constraints as the more recent GSWLC-2 \citep{{2018ApJ...859...11S}}, in order to stay with the more common SED fitting methodology. The principle difference between the SED fitting employed here and in S16 regards the parameterization of the star formation history. Namely, prior to running the final fits, we tested the performance of the SED fitting in order to find the parameters and variables that would potentially improve the quality of fitting over S16, judging by the average $\chi^2$. We took the first 1000 galaxies from GSWLC-D and performed the fitting with different SFH parameter combinations (changing the parameterizations of the of the old burst, as well as the ages and decline times of the young burst) and we found that some improvements were possible, as explained below. Tests were kept at 1000 galaxies in order to reduce the computation time over many trial runs. This number is entirely sufficient to reveal if one set of parameters performs better than another.

S16 uses a SFH parameterized as a double exponential, where the first (older) exponential commences at a fixed age (10 Gyr before the epoch of observation) whereas the second component is essentially flat in the SFR, but varies in duration (age) and intensity. In contrast, for this study we used the delayed exponential model from Appendix E of \citet{2019A&A...622A.103B}, but with the same age as the old exponential component used previously (10 Gyr) and without the extremely short decline models of 1 Myr. To this smooth delayed exponential we add the second, exponentially declining, component, again after the Appendix E. Unlike the second component in S16, this one includes various short decline rates and ages. We find that the new SF parameterization has somewhat smaller $\chi^2$ values for the best fitting models compared to S16, although the improvement is not dramatic.

For the fitting, the observed errors were added in quadrature with a calibration error ranging from 1-2\% for SDSS bands and 3-5\% for GALEX (S16). These errors were used instead of the 10\% calibration errors assumed by default in CIGALE. Finally, the redshift used in the fitting of composited was taken to be the average redshift of the two galaxies.

\begin{figure*}
    \includegraphics[width=0.33\textwidth]{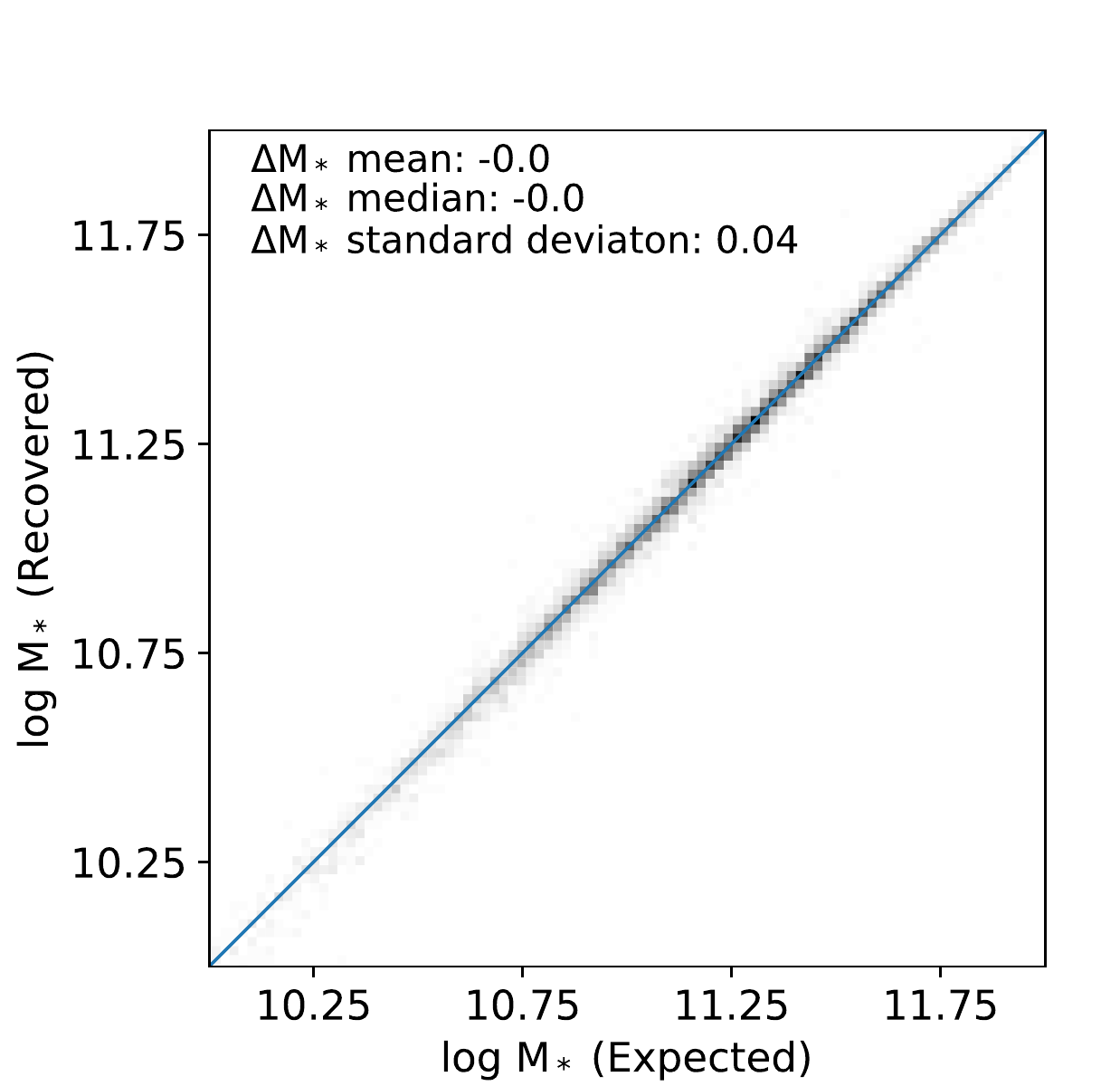}
    \includegraphics[width=0.315\textwidth]{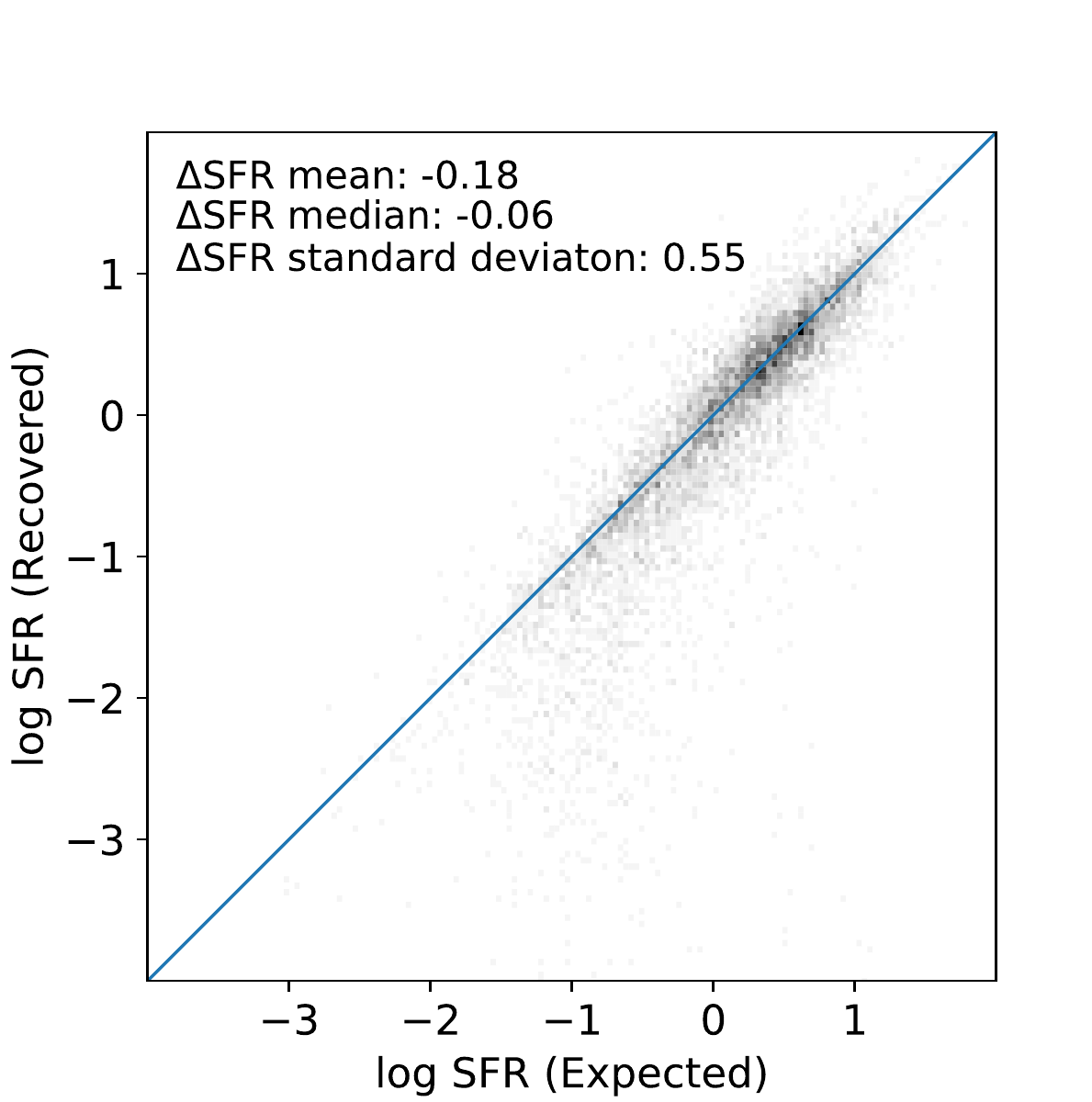}
    \includegraphics[width=0.32\textwidth]{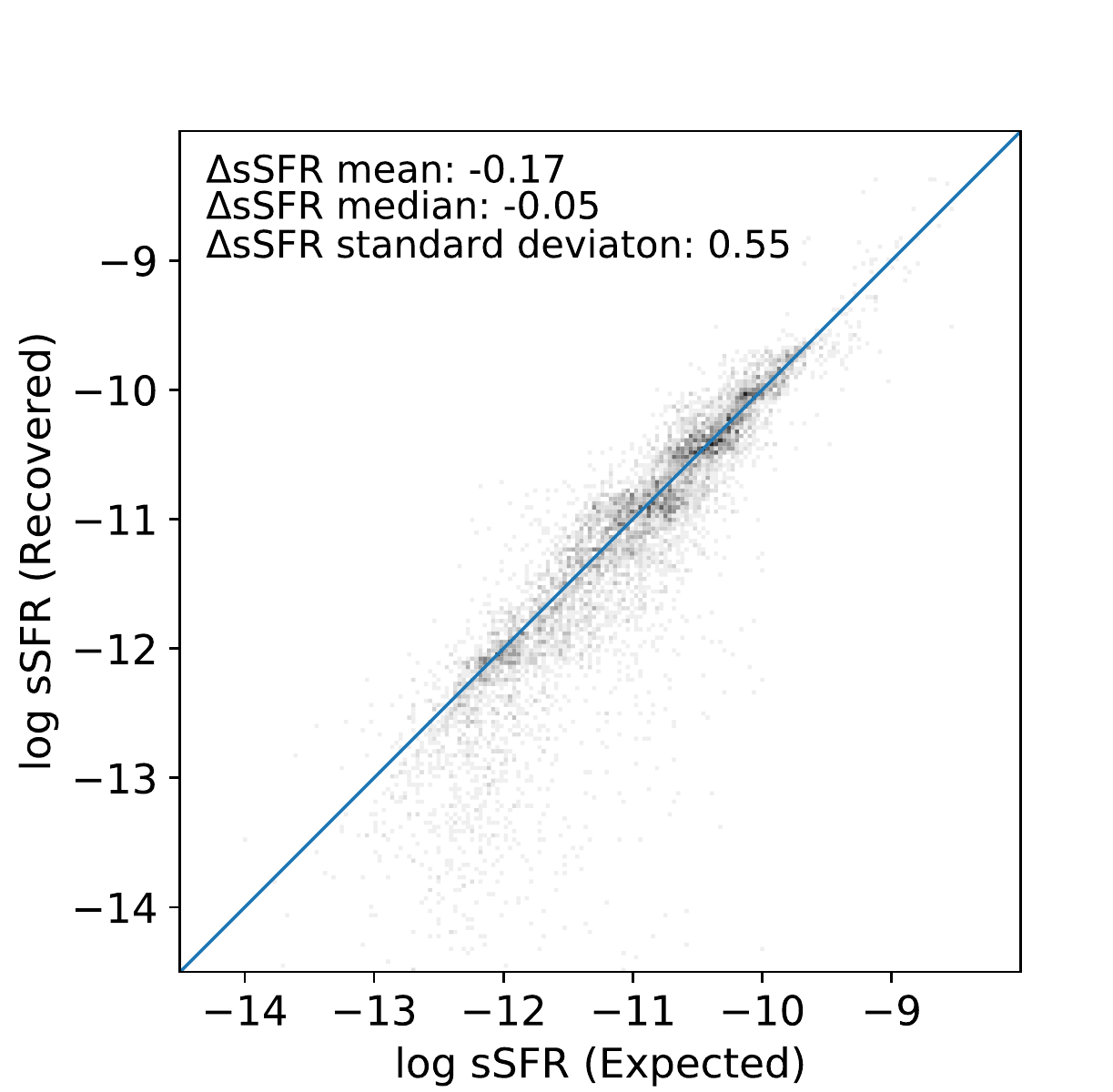}
    \caption{Comparison of the expected (sum of individual galaxies in a pair) versus recovered values (from the SED fit of the composite galaxy) of stellar mass, star formation rate, and specific star formation rate. Only pairs that have the stellar mass ratio corresponding to major mergers are shown.\label{fig:comparisons}}
\end{figure*}

\section{Results} \label{sec:results}
\subsection{Physical properties of parent sample and galaxy pair sample}
We first explore how the properties of galaxies in pairs before combining their light compare to those of the composite galaxies (after their SEDs were combined). This is done for the star formation rates (SFR), the specific star formation rates (sSFR), and the \emph{V}-band attenuation and is shown in Figure \ref{fig:1D_hist}. For the stellar mass the distribution shifts by upwards some 0.5 dex as expected because most of the massive galaxies are paired with another similarly massive galaxy (Figure \ref{fig:mass_ratio}). For SFR values, the values obtained for the composite galaxies have a pronounced high-SFR peak, whereas the low SFR present in the original distribution is suppressed. We confirm that this is not the result of there being fewer composites where both galaxies are quiescent (an equivalent of dry merging), but rather that the low SFR galaxies are overwhelmingly merged to higher SFR galaxies. For the sSFR, the composite values have a peak between the two peaks for galaxies prior to combining, which is expected given that many of the galaxy pairs are a combination of a passive plus star forming galaxy, so the result of combining would move the galaxy toward the green valley. Indeed it has been proposed that many of the galaxies in the green valley are the result of such mergers \citep{2009MNRAS.394.1713K}.The two histograms for the \emph{V}-band attenuation look similar and don't have much of an offset.

The stellar mass ratios of the galaxies making the composites are shown in Figure \ref{fig:mass_ratio}. The majority of mass ratios correspond to what would be the major mergers, which is expected because most of the pairs are at redshifts where only the massive galaxies ($\log M_*>10$) are detectable in the SDSS, so the difference between the galaxies in a given pair cannot be very high. Having a large fraction of composites pertain to the major mergers is a welcome aspect of this sample because it is the major mergers in which any biases due to the compositing of SFHs or dust properties would be accentuated. 

\begin{figure*}
    \centering
    \includegraphics[width=0.4\textwidth]{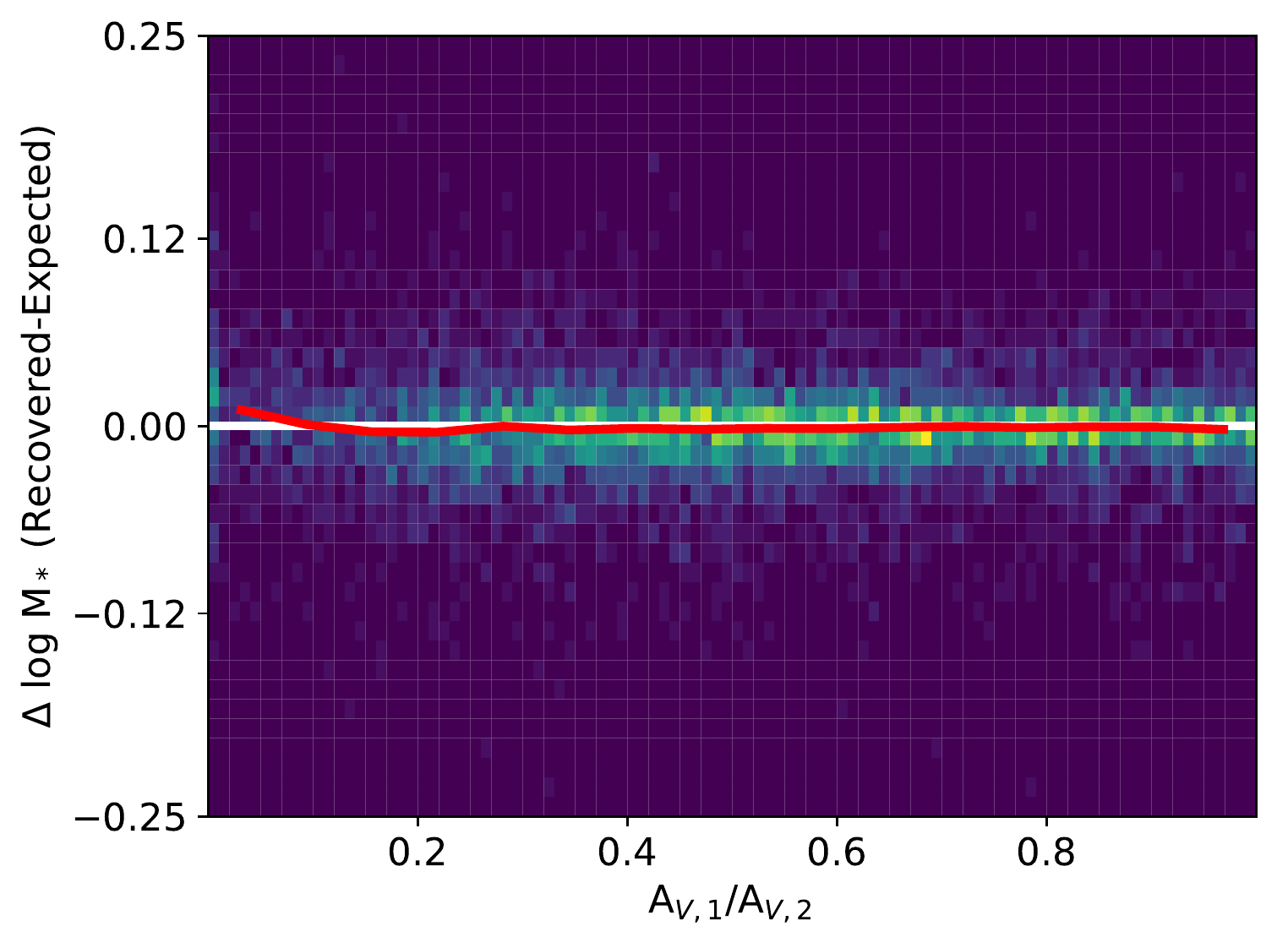}
    \includegraphics[width=0.4\textwidth]{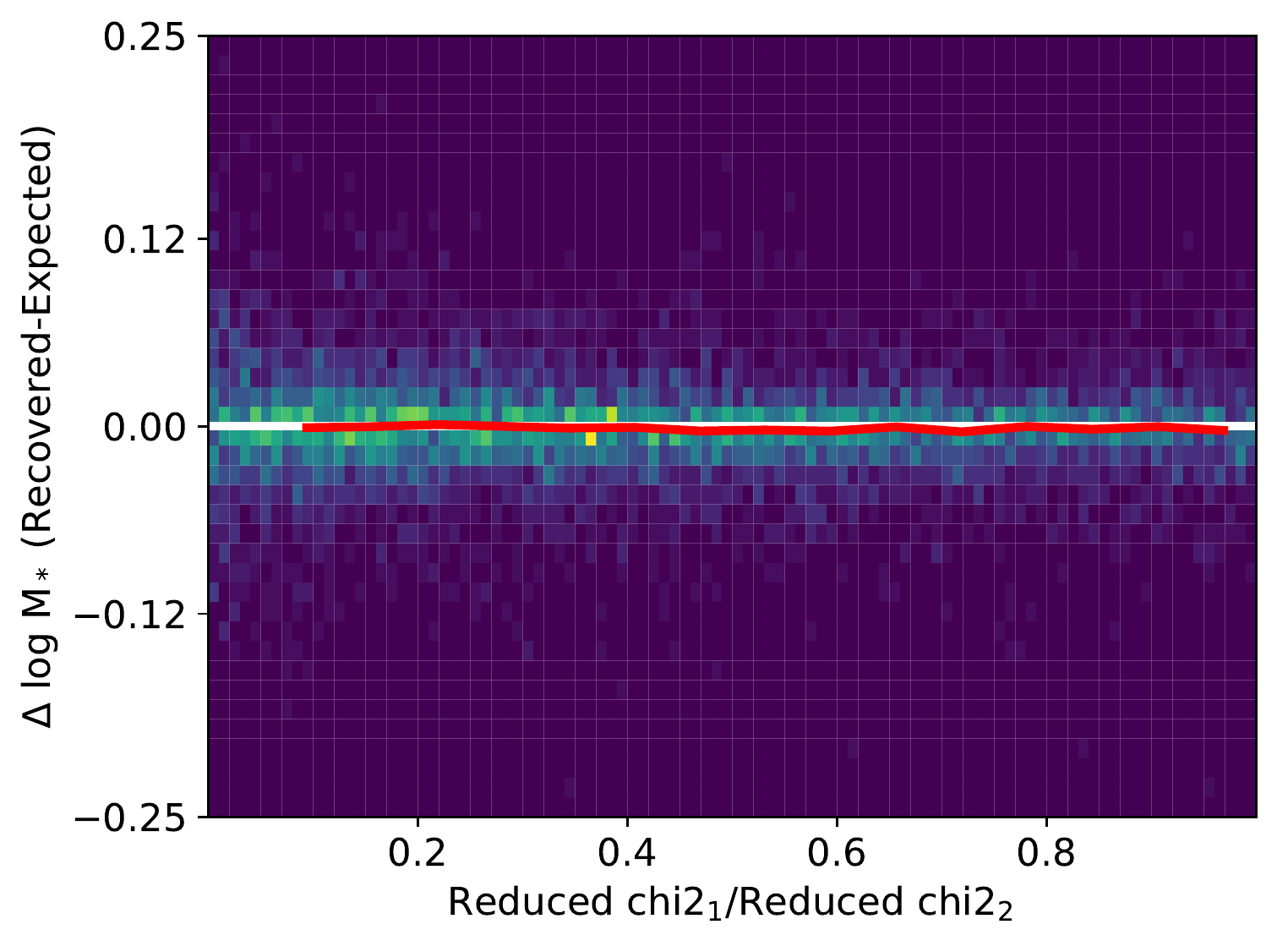}\\
    \includegraphics[width=0.4\textwidth]{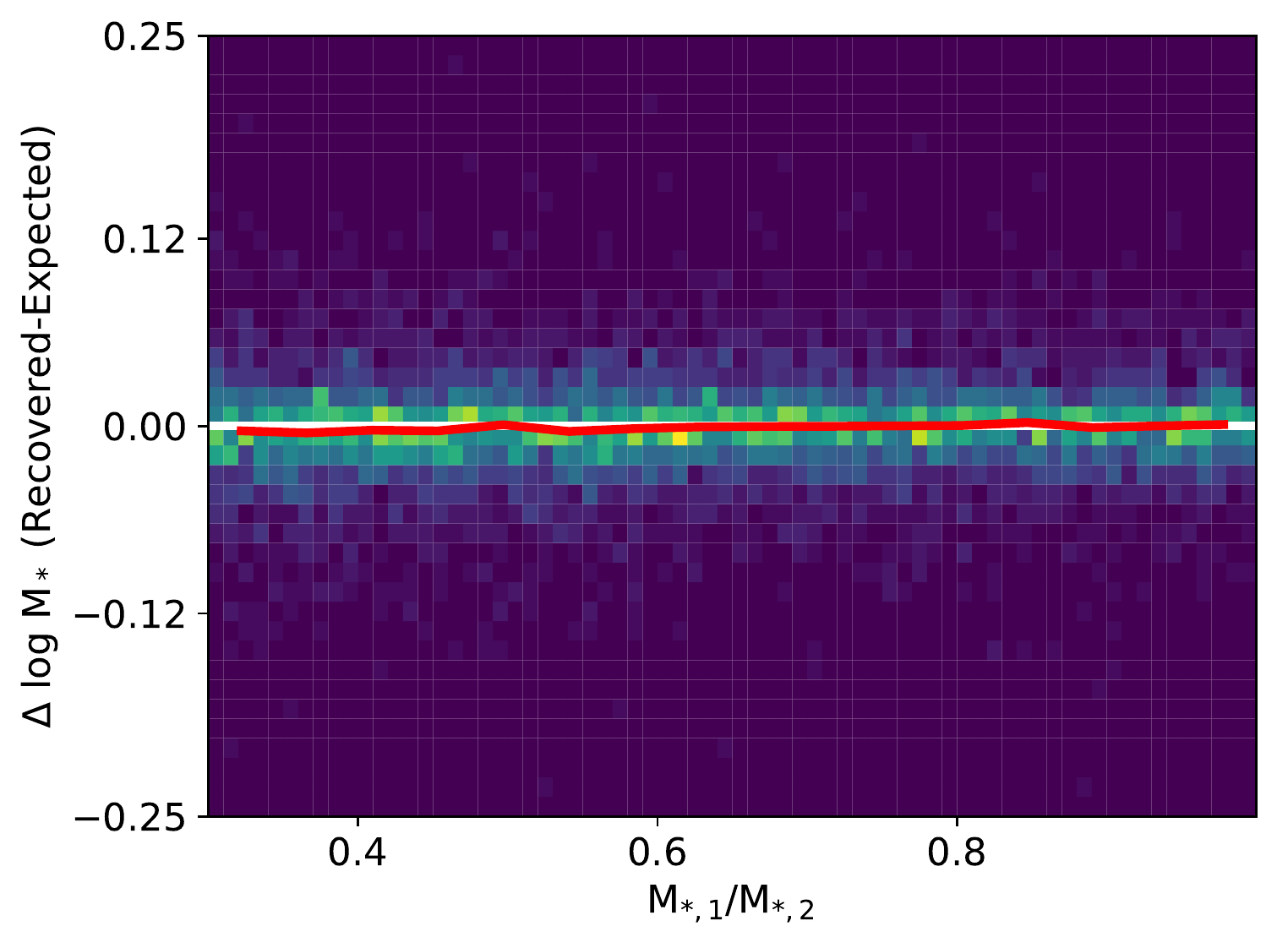}
    \includegraphics[width=0.4\textwidth]{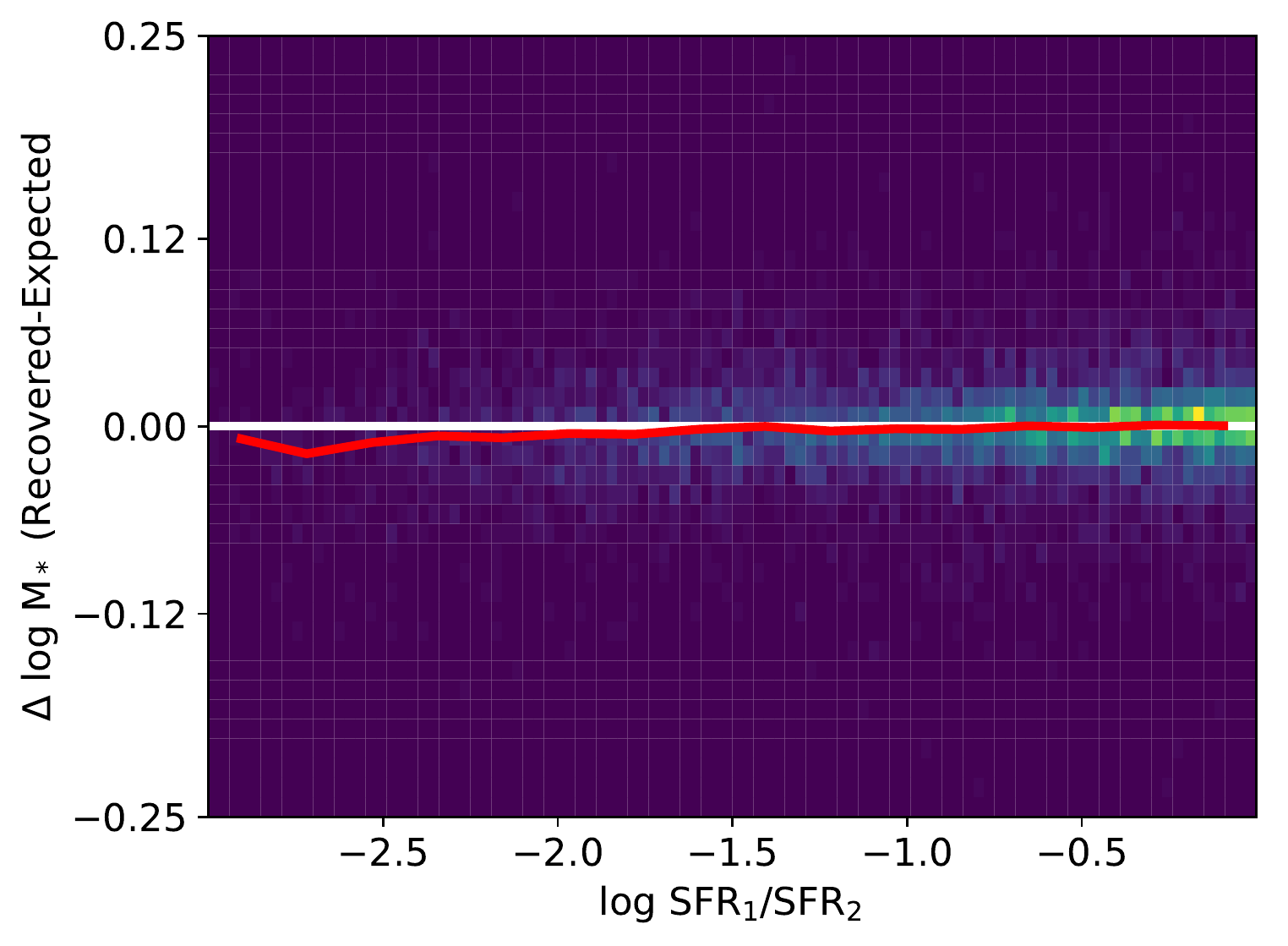}
    \caption{Difference in the recovered and expected stellar mass (the residuals) as a function of the contrast between the members in the pair, in \emph{V}-band attenuation, stellar mass, reduced $\chi^2$, and star formation rate. The white line is zero residual and the red line is the median value of the residuals. Only the pairs with the stellar mass ratios corresponding to major mergers are shown. \label{fig:results_hist_mass}}
\end{figure*}

The plot of the stellar mass versus the specific star formation rate was (Figure \ref{fig:M_SSFR_hist}) shows all 48,401 galaxies from the parent sample. On top of that are points for the merged galaxy pairs. The merged pairs have fewer low mass galaxies than the parent sample (low mass galaxies live in underdense environments and are unlikely to merge) and have a tail to lower specific star formation rates. We will see subsequently that some of these lower values are the result of a bias that leads to the underestimation of galaxies with already low sSFRs.

\subsection{Biases in parameters of composite galaxies}
In order to explore potential biases in parameters recovered for the composite galaxies, expected and recovered values were compared to see how well SED fitting on the composite galaxies agreed with the expected values. The expected values (``ground truth" in this comparison) were obtained by adding the parameter values for the individual galaxies, except in the case of the sSFR, where the sum of the SFRs divided by the sum of stellar masses is used. Figure \ref{fig:comparisons} shows comparisons of the expected versus recovered stellar masses, star formation rates, and specific star formation rates. We include only the composites ($M_{*,1}/M_{*,2} >0.3$). The plot of the mass comparison shows that the data points follow the one to one line very well (to within less than 0.01 dex), demonstrating that there is no overall bias in the stellar mass. There is also very little scatter (0.04 dex), comparable to, or even smaller than, the formal error in individual masses as estimated from the width of the probability distribution functions. For the star formation rate comparison, there does not seem to be a bias for log SFR$>0$ (median difference of only 0.01 dex), but there is a tail around log SFR (expected)$=-1$, in the sense that the recovered SFRs of these low-SFR galaxies are sometimes lower.
The specific star formation rate comparison is also mostly along the one to one line but there is a tail around log sSFR (expected) $=-12.5$ where again the recovered values are somewhat lower than expected. In order to find out what may be driving the biases for some of the low (s)SFR galaxies, we compared the estimated dust attenuation for the composite and of the original galaxy having the larger SFR (and therefore presumable carrying greater weight in the composite galaxy). Although there is a significant degree of scatter between the two dust attenuations, there are no apparent biases that would suggest that the biased dust estimate is behind the tail. We point out that the galaxies in the -13$<$log sSFR$<$-12 range are essentially passive, and their formal sSFRs are difficult to constrain with any degree of accuracy or precision, with the nominal errors in sSFR being on the order of the offset seen here between the recovered and the expected value.

\begin{figure*}
    \centering
    \includegraphics[width=0.4\textwidth]{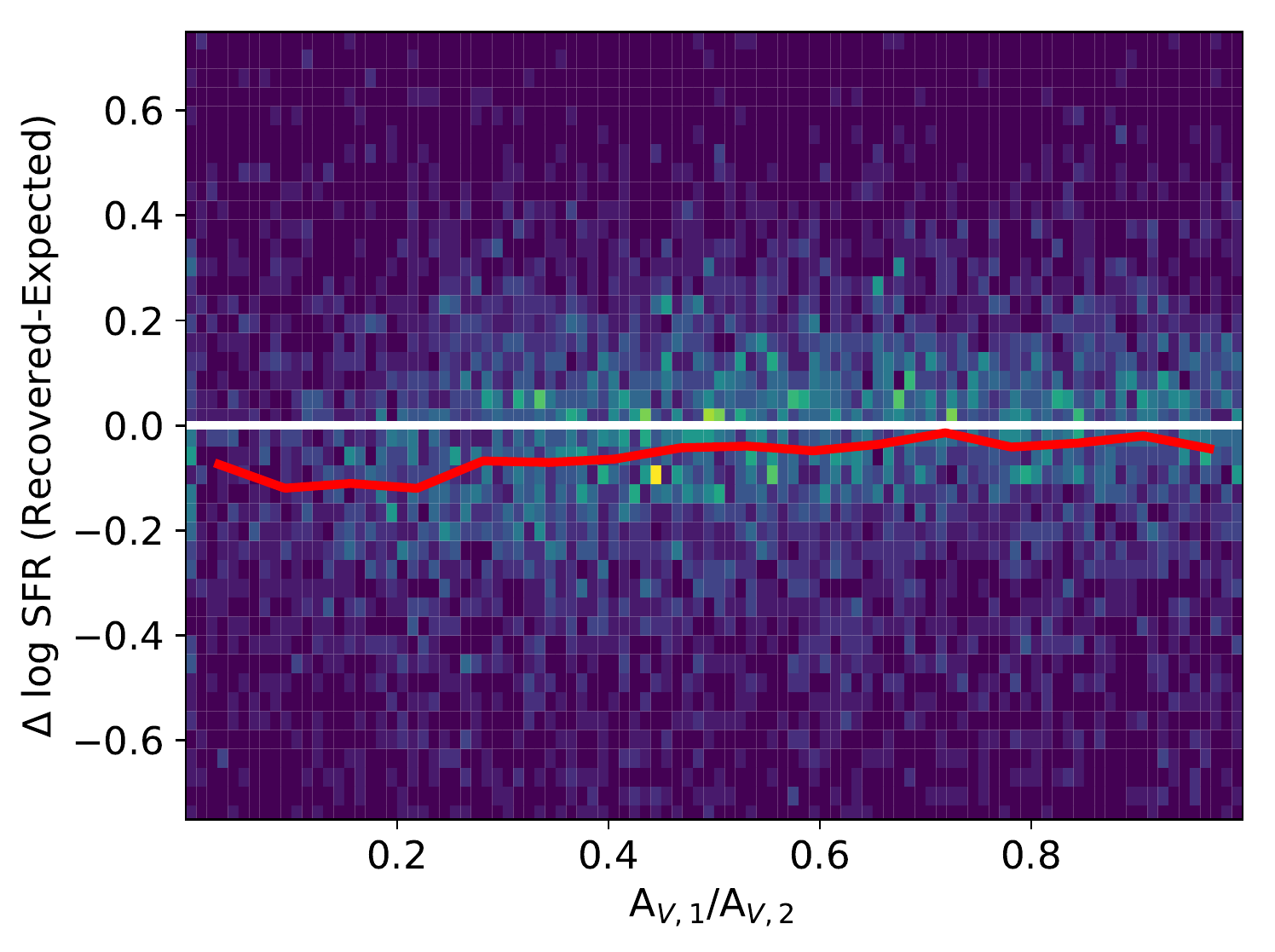}
    \includegraphics[width=0.4\textwidth]{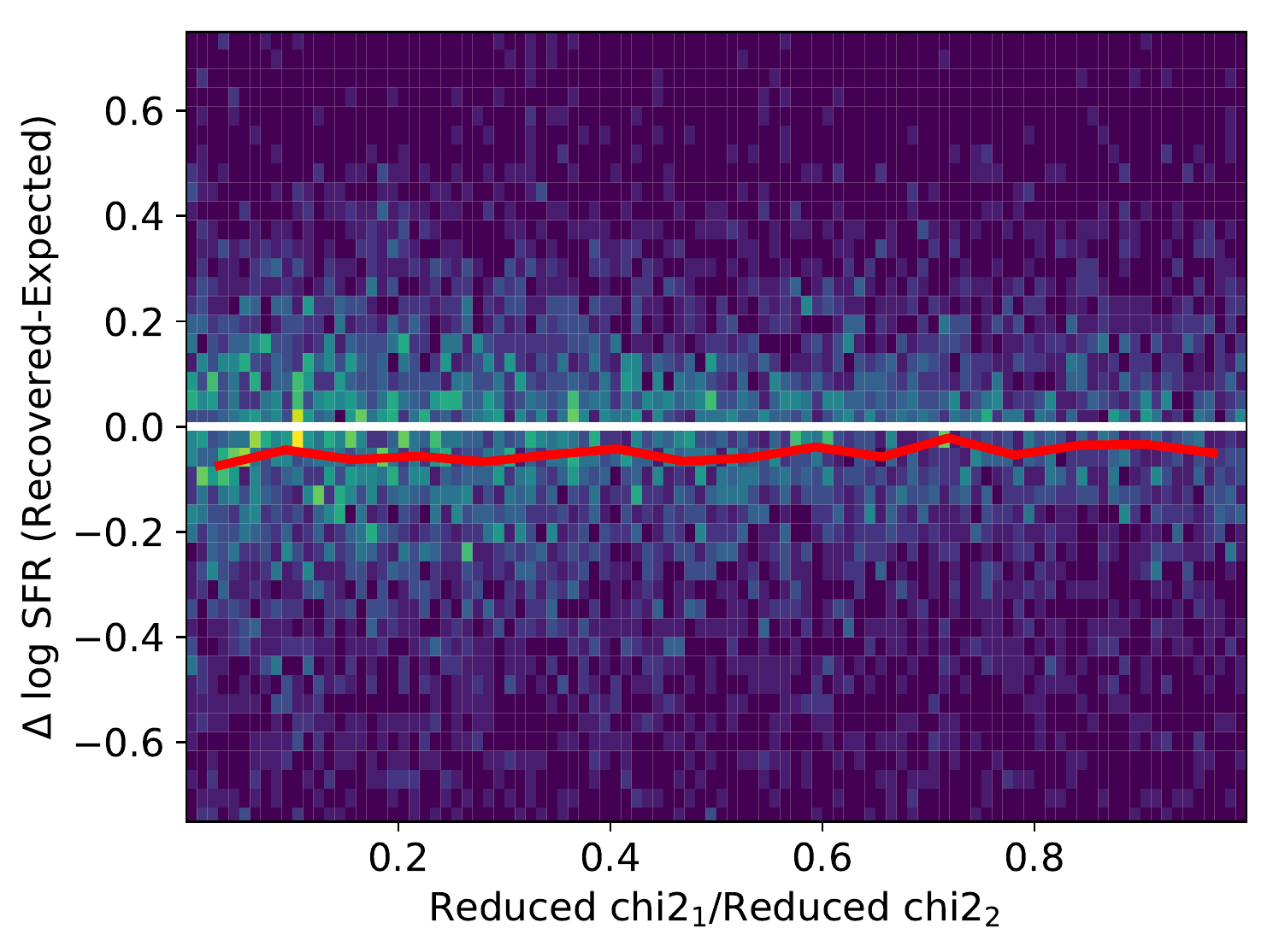}\\
    \includegraphics[width=0.4\textwidth]{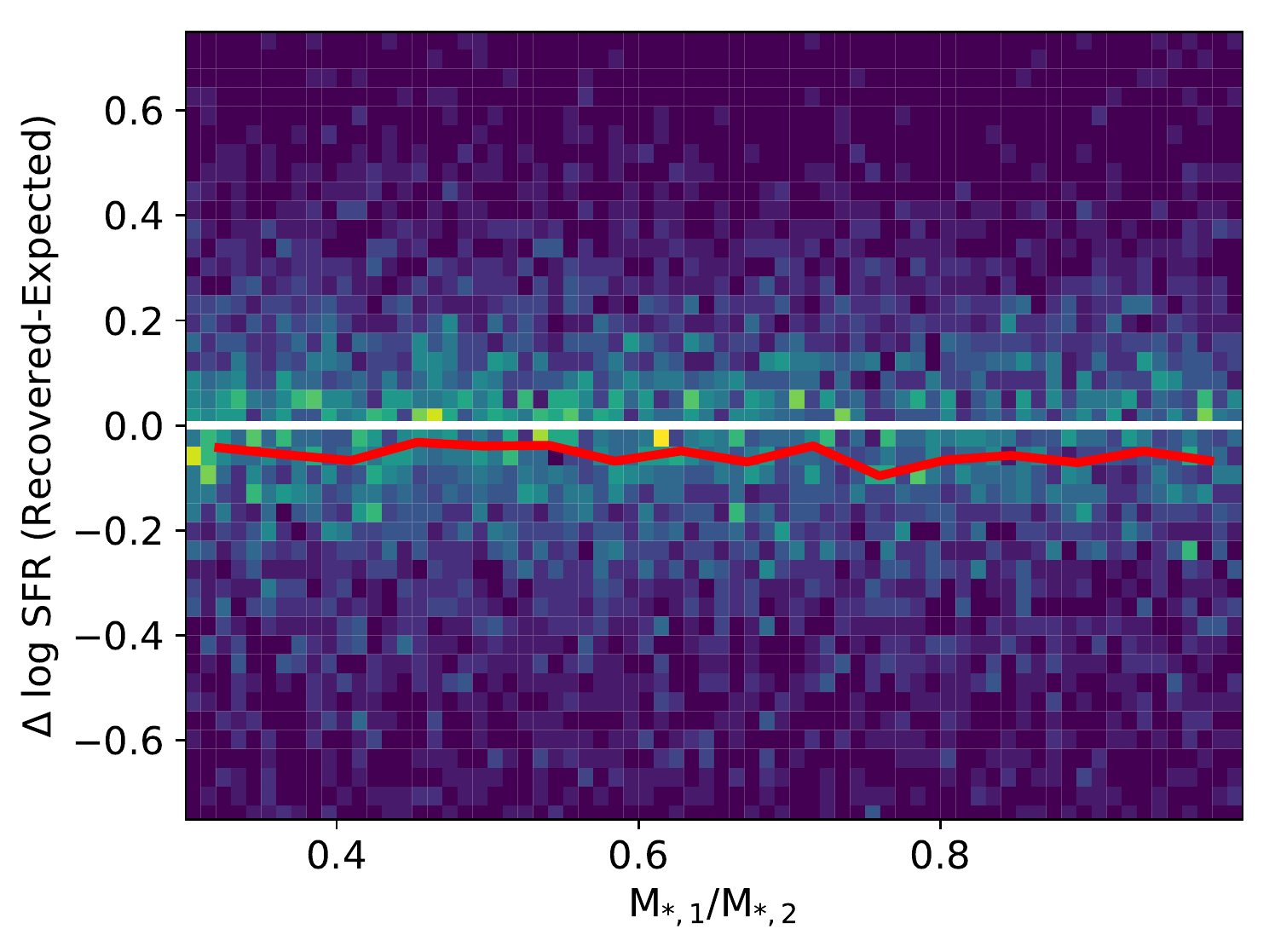}
    \includegraphics[width=0.4\textwidth]{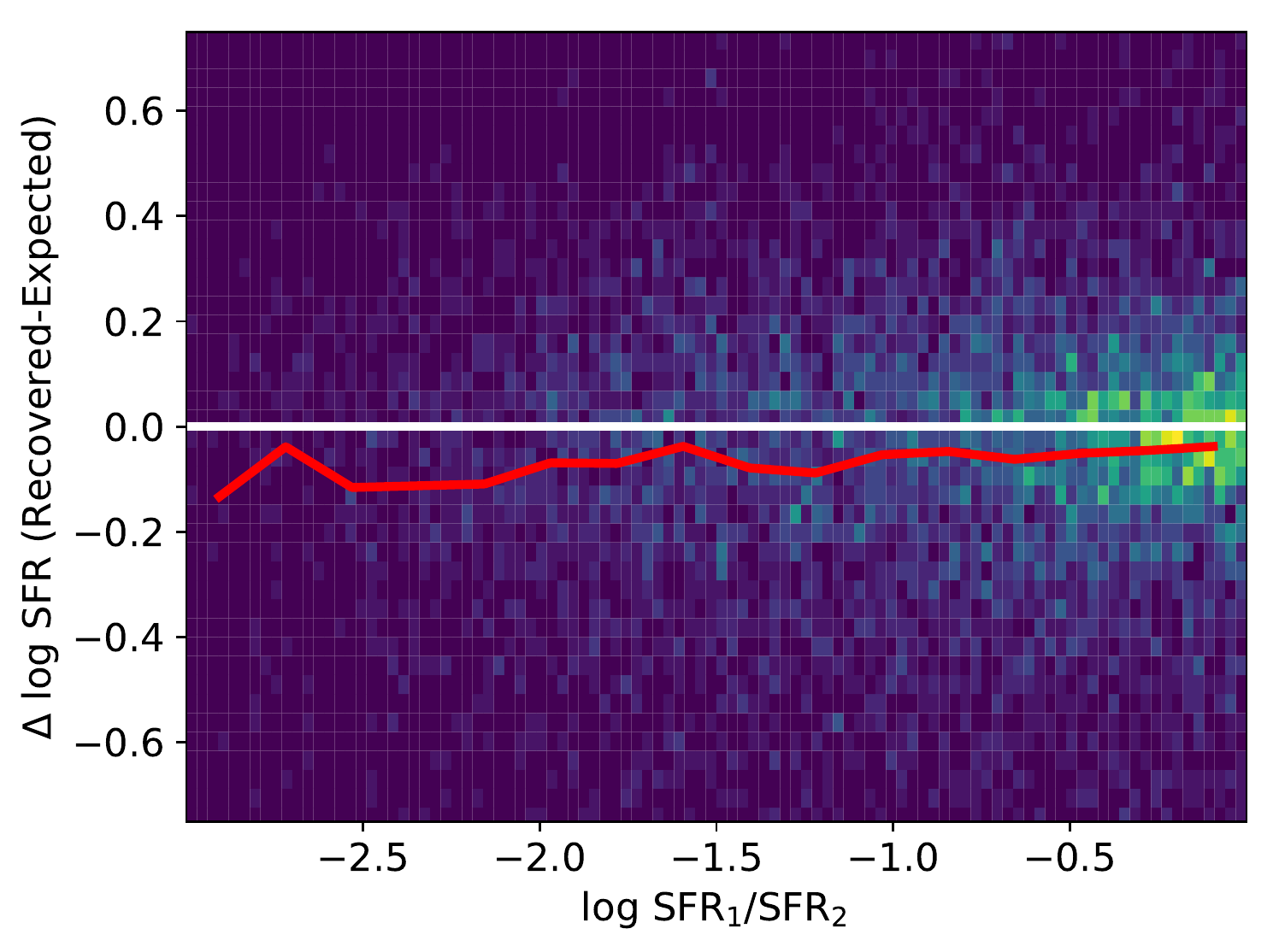}
    \caption{Difference in the recovered and expected SFR for (the residuals) as a function of the contrast between the members in the pair, in \emph{V}-band attenuation, stellar mass, reduced $\chi^2$, and star formation rate. The white line is zero residual and the red line is the median value of the residuals. Only the pairs with the stellar mass ratios corresponding to major mergers are shown. \label{fig:results_hist_SFR}}
\end{figure*}

Whereas the direct comparison does not reveal any major biases (except for a small number of low SFR and sSFR galaxies), it is still of interest to explore if any residuals may emerge for particular subsets of composite galaxies. For example, one might imagine that the biases increase for pairs that have more comparable mass ratios. In Figure \ref{fig:results_hist_mass} we show stellar mass residuals between expected and recovered mass as a function of the contrast in: \emph{V}-band attenuation, the quality of the fit (reduced $\chi^2$ value), the stellar mass, and the SFR. The median value of the residuals is shown by a red line. Residuals stay very close to zero (white line) in all four plots, showing that there are no systematic stellar mass residuals even for cases where one galaxy in the pair has a very different dust attenuation from the other, where composites involve nearly equal mass galaxies, or where one fit is much worse than the other.

Figure \ref{fig:results_hist_SFR} shows the residuals as a function of the contrast between the same parameters used in Figure \ref{fig:results_hist_mass}, but now for SFR. The red line again shows the median value of the residuals. We see that the residuals are not driven by increasing mass contrast or the difference in the quality of the fits (upper right and lower left panels), but they do increase somewhat when one galaxy has much less dust than another or when a passive and active star forming galaxy are combined. However, even in those cases the residuals are around 0.1 dex, which is comparable to or smaller than a typical SFR random error (0.1 to 0.6 dex, Figure 6 in S16).

\section{Discussion and Conclusions} \label{sec:discussion}
Merging can cause dramatic changes to a galaxy's star formation \citep{{2011EAS....51..107B}}, and this discontinuity in SFH may pose difficulties for the accurate determination of SFR, especially if the model SFHs lack a recent burst component. Leaving those issues aside, our study focuses on the biases that may emerge even just from the compositing of the light of two galaxies, as in the case of a post-merger system. While no previous study has examined this problem, the character of our study is most similar to that of \citet{{2015MNRAS.452..235S}}, who compared stellar mass estimates obtained in the UV/optical SED fitting by treating the galaxies as unresolved and also as the sum of pixels. Their sample consisted of 67 very nearby galaxies (mean redshift of 0.006). They found that for galaxies with higher sSFR the mass from the integrated SED was lower than the one from summed pixels, which was attributed to biases arising from younger stars outshining older ones. It needs to be pointed out that for a typical star-forming galaxy on the main sequence (log sSFR$\sim$10), they find the bias of just 9\%, or 0.04 dex. It reaches 0.11 dex only for very strong starbursts (log sSFR=-8), which are rare in the local universe. Our comparison of composite vs. individual masses shows no stellar mass biases greater than 0.01 dex (Figure \ref{fig:results_hist_mass} mass panels), seemingly at odds with \citet{{2015MNRAS.452..235S}}. The probable reason for that is that \citet{{2015MNRAS.452..235S}} contrast pixel-based estimates to fully unresolved ones, which will maximize any differences, whereas we compare the sum of two already unresolved galaxies to their composite. Indeed, they find that if their comparison is made between the images degraded to 3 kpc (rather than pixel-level resolution of $\sim$0.1 kpc) and the fully unresolved one, the systematic differences between the two are no longer present. \citet{{2015MNRAS.452..235S}} investigate the effects of the resolution only for $M_*$, so we cannot compare our results for the SFR.

As shown in Section \ref{sec:results}, no major systematics were found in the determination of the principle galaxy properties, the light of which is the composite of two galaxy SEDs. In Figure \ref{fig:comparisons}, the recovered stellar masses match the expected values well, with median differences within 0.01 dex and small standard deviations. For SFRs, no biases are present for galaxies on the star forming main sequence (median differences within 0.01 dex). sSFR comparisons show a bias around log sSFR (expected) $=-12.5$, which we attribute to the general difficulties involved in getting the accurate SFRs for practically quiescent galaxies. The mass residuals do not depend on the contrast of the  \emph{V}-band attenuation, reduced $\chi^2$ value, stellar mass, and SFR, whereas the SFR residuals are somewhat sensitive to extreme dust and SFR contrast. 

Overall, regardless of the different dust properties and SFH, the parameters of the composite do not suffer from significant biases in the SED fitting. From this we conclude that despite the great complexity that the SFHs may exhibit in reality due to the different processes that boost and suppress star formation (e.g. \citet{{2016MNRAS.457.2790T}}), when it comes to the ability to derive physical parameters, even relatively simple two-component prescriptions are sufficient to obtain accurate results. This conclusion has a practical significance for SED modelers, but also informs us that parametric models of SFH still have their merits in the high-level understanding of galaxy evolution.

Another important methodological contribution of this study for SED modelling is the simplicity of the approach of combining fluxes to study potential systematics. This approach deserves to be incorporated into a suite of robustness diagnostics, such as the covariance analysis (corner diagrams, \citet{{2017ApJ...837..170L}}), as well as mock SED fitting (SED fitting to best-fitting photometry obtained from the SED fitting itself, \citet{{2019A&A...622A.103B}}, \citet{{2009ApJ...700..161S}}).

Only the main physical parameters were explored in this study and there may be other parameters which would be more affected by the biases in the SED fitting in cases of combining the flux of galaxies. On the other hand, here we explore just the more usual SED fitting involving stellar emission. The inclusion of dust emission in the infrared, as is done in the so called ``energy balance" SED fitting, may help further reduce any systematics \citep{{2008MNRAS.388.1595D}, {2018ApJ...859...11S}}.

\begin{acknowledgements} The construction of GSWLC was funded through NASA awards NNX12AE06G and 80NSSc20K0440.
\end{acknowledgements}

\end{document}